# Characteristics of superconducting tungsten silicide $W_xSi_{1-x}$ for single photon detection

X. Zhang, A. Engel, Q. Wang, and A. Schilling

*Physics-Institute, University of Zürich, Winterthurerstrasse 190, 8057 Zürich, Switzerland*

A. Semenov, M. Sidorova, and H.-W. Hübers

*DLR Institute of Optical Systems, Rutherfordstrasse 2, 12489 Berlin, Germany*

I. Charaev, K. Ilin, and M. Siegel

*Institute of Micro- und Nano-electronic Systems, Karlsruhe Institute of Technology (KIT), Hertzstrasse 16, D-76187 Karlsruhe, Germany*

Superconducting properties of three series of amorphous $W_xSi_{1-x}$ films with different thickness and stoichiometry were investigated by dc transport measurements in a magnetic field up to 9 T. These amorphous $W_xSi_{1-x}$ films were deposited by magnetron co-sputtering of the elemental source targets onto silicon substrates at room temperature and patterned in the form of bridges by optical lithography and reactive ion etching. Analysis of the data on magnetoconductivity allowed us to extract the critical temperatures, superconducting coherence lengths, magnetic penetration depths, and diffusion constants of electrons in the normal state, as functions of film thickness for each stoichiometry. Two basic time constants were derived from transport and time-resolving measurements. A dynamic process of the formation of a hot-spot was analyzed in the framework of a diffusion-based vortex-entry model. We used a two stage diffusion approach and defined a hotspot size by assuming that the quasiparticles and normal-state electrons have the same diffusion constant. With this definition and these measured material parameters, the hot-spot in the 5 nm thick $W_{0.85}Si_{0.15}$ film has a diameter of 107 nm at the peak of the number of non-equilibrium quasiparticles.

I. Introduction

During the past decade, superconducting nanowire single-photon detectors (namely SNSPDs or SSPDs)[1] have been intensively studied because of their shorter timing jitter, much higher detection efficiency and nearly negligible dark-count rate compared to other single-photon technologies in the near infrared spectral range.[2-5] Great efforts have been made to improve the detection efficiency for visible and infrared photons. NbN films were deposited onto different substrates under varying conditions in order to find the most suitable NbN film parameters.[6-8] For detectors with optimized stoichiometry[9] the superconducting transition temperature $T_c$ varies from 10 to 11 K. Other Nb-based materials have also been considered.[10] NbTiN has a slightly smaller $T_c$ as compared to NbN ($T_c$ of NbTiN is 1 K lower than $T_c$ of NbN) at film thickness $d < 20$ nm, but it does not require such high substrate temperature to grow epitaxial films (details of the dependence of the transition temperature in NbTiN on the film thickness and growth conditions can be found in reference 11). On the other hand, devices made from NbSi have a much lower $T_c \approx 2$ K, which results in a higher relative detection efficiency for low-energy photons.[12] Similar results can be achieved with devices from TaN which have intermediate $T_c$ values from 6 to 10 K and, consequently, the advantage of still being able to operate efficiently between 2 and 4 K.[13-16] Recent results obtained with detectors from amorphous superconducting films MoGe,[17] WSi,[18-21] and MoSi,[22] which all have $T_c$ in the range from 5 to 7.5 K, retain the promise of significant improvement in both detection efficiency and spectral range extending the sensitivity further into the infrared. So far, the highest detection efficiency (DE) reported for SNSPD of the order of 93% has been achieved with WSi.[20]

A very important limiting factor of the intrinsic detection efficiency are the constrictions in the devices.[23] Non-uniformities of the superconducting film itself or local imperfections within the nanowire, which are introduced during the structuring process, can result in reduction of the local density of the critical current. This effect becomes more pronounced for narrower nanowires. It has been also shown that it can severely reduce the internal quantum efficiency.[24, 25] Amorphous thin films are generally much more homogeneous at the relevant length scale from a few nanometers up to a few tens of nanometers as compared to the epitaxial films. But this alone cannot explain the extraordinary performance of WSi-SNSPD. Intuitively, an incident photon with a given energy should break more Cooper pairs in a film with smaller superconducting energy gap and thus the resulting detection efficiency should be higher. Indeed, materials with smaller superconducting energy gap do extend their spectral range of 100% intrinsic detection efficiency towards larger wavelengths. However, detection efficiency of the wire-structure (device detection efficiency) remains relatively low and is limited to its absorbance. The latter can be driven to 100% by applying multilayer resonators and improving optical coupling. Differently, improvements of the device detection efficiency can be achieved by optimization of either geometry or material of the wire-structure. Besides the wire geometry and the superconducting energy gap, other factors restricting the detector performance may not have been clearly identified. An indication that material parameters are indeed important for the detection mechanism and the performance of SNSPD has come from a comparison between NbN and NbC.[26]

In order to understand the drastic increase of the spectral roll-off in the detection efficiency of WSi-SNSPD, we need to have a detailed knowledge about properties of WSi films beyond their superconducting transition temperature. To achieve this, we studied the electronic transport parameters of $W_xSi_{1-x}$ films with different thicknesses. Additionally, the time constants of the $W_xSi_{1-x}$ films that are relevant for photon detection were acquired by means of measurements of the magnetoconductivity and time-resolving recovery after femtosecond optical excitation. We used our diffusion-based vortex-entry model[27-29] for analyzing the detection mechanism and defining the size of the hot-spot (namely the diameter of the hot-spot) which is produced in our films by single-photon excitation.

II. Sample preparation and experimental approach

Superconducting amorphous $W_{0.75}Si_{0.25}$, $W_{0.8}Si_{0.2}$ and $W_{0.85}Si_{0.15}$ films were grown on silicon substrates by DC magnetron sputtering of a pure W (99.95%) target and RF magnetron sputtering of a pure Si (99.999%) target in argon (Ar) atmosphere, at a total pressure of 3 mTorr. The nominal thickness $d$ of the resulting films was inferred from the predetermined growth rate and the deposition time. The $W_xSi_{1-x}$ films were protected from oxidation by a silicon capping layer with a thickness of 1.5 nm. We deposited several series of films with thickness of 5, 10, 20, 50 and 100 nm for each stoichiometry. The strips were patterned from the freshly deposited films by optical lithography and reactive ion etching with the strip width ranging from 10 to 200 μm. Each strip had six contacts, with 4 pads used for resistivity measurements. We measured the strip sizes (width and length) through the inspections with a scanning electron microscope (SEM). All the calculations and data extraction are based on the SEM measurements and the nominal film thicknesses. SEM images of the strip pattern are shown in Fig. 1(a). We use a wedge wire-bonder for electrical connections of our WSi samples in a four-point probing configuration. The resistivity measurements were carried out in a physical property measurement system (PPMS) from *Quantum Design* under various magnetic fields up to 9 T. For measurement in a magnetic field, the field was directed perpendicular to the microstrip surface.

In order to trace the recovery of superconductivity (namely the relaxation time of hot-spot) in the time-resolving experiment, we adopted a bow-tie microbridge. The microscopic images of the bow-tie and of the whole structure are presented in Fig. 1(b). The fabrication process of the microbridge includes three lithographic steps. Firstly, two small pads (see the inset in Fig. 1(b)) were patterned onto the WSi film by means of the electron-beam lithography. These two pads were separated by a slit, which defines the length $L$ of the future microbridge (see the enlarged part in Fig. 1(b)). PMMA resist with a thickness of 150 nm on top of WSi film was exposed using 10 kV electron beam with a dose of 120 μC/cm$^2$. The Nb / Au bi-layer consisting of 8 nm Nb and 100 nm Au was deposited on top of the WSi film by magnetron sputtering at a partial pressure of $P_{Ar} = 5 \times 10^{-3}$ mbar. The lift-off process was carried out in a warm acetone and ultrasonic bath. To pattern the large contact pads, the substrate was covered by photo-resist with a thickness 950 nm. By the subsequent photolithography, magnetron sputtering and lift-off process, a three-layer Nb/Au/Nb (8 / 250 / 15 nm) sandwich was formed on the surface of the substrate. The width of bridge, $W$, was defined by the e-beam lithography over negative resist. Finally, the WSi microbridge was etched with Ar ion milling. During the etching process the upper Nb layer of the large contact pads protected the gold layer from Ar ions. The dimensions ($L$ and

*W*) of the microbridge in the slit of the bow-tie and the embedding co-planar transmission line were designed in such a way that the microbridge in the normal-state and the line both had an impedance of approximately 50 Ω (see the inset in Fig.1 (b)). While widths of all microbridges stayed the same (5 µm), the length varied between 700 and 900 nm. In the time-resolving experiment, the beam of a femtosecond pulse laser with a wavelength of 800 nm was positioned over the center of the bow-tie. The beam diameter at the bow-tie was much larger than both *L* and *W* that ensured uniform excitation of the microbridge. The electric response to laser pulses was monitored with a time-resolving (resolution 1.25 ps) readout.

III. Experimental data

We characterized a series of $W_xSi_{1-x}$ films to obtain the fundamental material parameters such as superconducting coherence length $\xi(0)$, normal-state electron diffusion constant $D_e$, electron density-of-state at the Fermi surface $N_0$, energy gap $\Delta$, magnetic penetration depth $\lambda(0)$, and characteristic time-scales $\tau_{qp}$ and $\tau_r$, which are relevant for the dynamic photon detection process. These transport parameters and the hot electron relaxation time $\tau_{qp}$ can be acquired from systematic transport measurements (see subsection A and B), while the recovery time of superconductivity $\tau_r$ can be obtained from the dynamic response after a photon absorption (see subsection C).

A. Resistivity

The square resistance at each temperature $R_s(T)$ was calculated from the measured total resistance and the strip geometry. In Fig. 2 it is shown for a thick (*d* = 100 nm) film and for a two dimensional (*d* = 5 nm) film. As the ambient temperature decreases, the film enters the region of the superconducting transition and the square resistance starts to decrease. The mean-field superconducting transition temperature $T_c$ can be estimated by taking into account the contributions to the total conductivity from fluctuating Cooper pairs.[30, 31] When expressed in terms of the measured square resistance, this contribution for three (3D) and two (2D) dimensional films takes the forms[32, 33]

$$R_s(T) = \frac{1}{\frac{1}{R_{ns}} + \frac{5}{32} \cdot \frac{e^2}{\hbar \xi(0)} \cdot d \cdot (\frac{T_c}{T-T_c})^{0.5}}, \qquad (1)$$

$$R_s(T) = \frac{1}{\frac{1}{R_{ns}} + \frac{3}{16} \cdot \frac{e^2}{\hbar} \cdot (\frac{T_c}{T-T_c})}. \qquad (2)$$

Here $e$ is the elementary charge; $\hbar$ is the reduced Planck constant; $\xi(0)$ is the coherence length; $d$ is the film thickness; and $R_{ns}$ is the normal-state square resistance. The fluctuation conductivity terms have included both Aslamazov-Larkin (AL) and Maki-Thompson (MT) fluctuations (here we have made simplifications to the MT term, and the detailed MT expression for 2D films can be found in Ref. 33). We achieved good description of the measured data with Eq. (1) for film thicknesses larger than 10 nm, and with Eq. (2) for the 5 nm and 10 nm thick films. The best fits are shown in the insets in Fig. 2(a) and (b). The fitted $T_c$ is consistent with a $R_{ns}(7K)/2$ criterion. In Fig. 2(c), the mean-field transition temperature is plotted as a function of the film thickness. Our data are similar to the results from other groups.[18, 34]

By measuring the superconducting transitions at different magnetic field, we obtained the magnetic field dependence of the transition temperature $T_c(B)$ using the $R_{ns}/2$ standard criterion. According to the Ginzburg-Landau theory, $T_c(B)$ should approximately be linear in $B$ at temperatures close enough to $T_c(0)$, as it is depicted in Fig. 3. The critical fields at zero temperature for each film can be obtained by extrapolating the line to its intercept with the $B$-axis. The thicker films show a larger slope and thus a larger critical field. The experimental data deviate from the linear dependence when the applied magnetic field is comparatively large or rather small. As a result, these extrapolated $B_{c2}(0)$ values are larger than the actual critical fields, and a more realistic value $B_{c2}(0)$ can be obtained by multiplying them with a factor of 0.69.[35, 36] In this paper, all the calculations are based on the linearly extrapolated $B_{c2}(0)$, therefore the calculated coherence length is the Ginzburg-Landau (GL) coherence length. From the GL theory, the zero-temperature critical magnetic field $B_{c2}(0)$ is related to the GL coherence length[35]

$$B_{c2}(0) = \frac{\Phi_0}{2\pi \xi^2(0)}, \qquad (3)$$

where $\Phi_0$ is the magnetic-flux quantum. With the decrease of the films thickness, the zero-temperature GL coherence length exhibits a significant increase, as it is shown in Fig. 4. In the amorphous $W_xSi_{1-x}$ superconducting films studied here, the Ginzburg-Landau coherence length is larger than the coherence length in traditional NbN materials for SNSPD fabrication (see Table I), which may make SNSPDs from $W_xSi_{1-x}$ material more robust against the inhomogeneities like local variations of the film thickness or constrictions within the nanowire. The diffusion constant of the normal-state electrons can be determined from the slope of the $T_c(B)$ curve as[37]

$$D_e = \frac{4k_B}{\pi e} \cdot \left(\frac{dB_{c2}}{dT}\right)^{-1}\bigg|_{T \to T_c(0)}, \tag{4}$$

where $k_B$ is the Boltzmann constant. The corresponding diffusion constants $D_e$, which were calculated by Eq. (4) for different films, are plotted in Fig. 5 as a function of the film thickness. There is no systematic variation of the diffusion constant with the stoichiometry. At the same time, the diffusion constants of the thinnest (5 nm) WSi films are about 10% larger than the diffusion constants of thicker (30 nm and more) films. It is expected that the normal-state electron in materials with a larger diffusivity will diffuse farther away from the photon absorption position within equal time intervals, which in turn will result in larger hot-spots.[38, 39] From the diffusion constant we can estimate the electronic density of states $N_0$ at the Fermi level via the Einstein relation[40, 41]

$$N_0 = \frac{4k_B}{e^2 \rho_n D_e}. \tag{5}$$

Here $\rho_n$ is the normal-state resistivity, which is calculated from the normal-state resistance $R_n$, strip sizes and the film thickness. It is interesting to note that the calculated electronic densities of states for our amorphous WSi films in table I are an order of magnitude higher than those in $W_xSi_{1-x}$ with crystalline structures ($N_0 = 3.64 \times 10^{46}$ m$^{-3}$J$^{-1}$ for $WSi_2$ and $N_0 = 1.36 \times 10^{46}$ m$^{-3}$J$^{-1}$ for $W_5Si_3$).[42, 43]

Since we have not directly measured the values for the superconducting energy gap for $W_xSi_{1-x}$ films, we used the BCS relation $\Delta(0) = (\pi/e^\gamma)k_BT_c$ with $\gamma=0.577$.[44] With the critical temperatures extracted from the fits in Fig. 2, we can calculate the superconducting gaps, and also obtain the magnetic penetration depths at zero temperature through[41]

$$\lambda(0) = \left(\frac{\hbar \rho_n}{\pi \mu_0 \Delta(0)}\right)^{0.5}, \tag{6}$$

where $\mu_0$ is the vacuum permeability. For each stoichiometry, the dependence of the magnetic penetration depth on the film thickness is shown in Fig. 6. The magnetic penetration depths increase with the reduction of the film thickness, especially for the ultrathin films, which are used for SNSPDs fabrication. All the calculated transport parameters presented here are summarized in Table I. For comparison, two groups of data for NbN and one group of data for TaN are also listed at the bottom of the same table.

B. Magnetoconductivity

A photon that is absorbed in the nanowire creates a highly excited electron which consequently diffuses along the nanowire. It subsequently loses its energy and thermalizes with a time scale $\tau_{qp}$ via inelastic scattering events, thereby breaking Cooper pairs and creating quasiparticles.[27-29] For a superconductor at a temperature near $T_c$, the inelastic scattering occurs due to electron-electron (*e-e*) interaction, electron-phonon (*e-ph*) interaction, and superconducting fluctuation (*e-fl*).[37, 45-47] At high temperatures, namely $T \gg T_c$, $\tau_{qp}$ is mainly determined by the *e-e* and the *e-ph* interactions, while at a temperature slightly above $T_c$, $\tau_{qp}$ is governed by fluctuations. This corresponding characteristic timing constant $\tau_{qp}$ can be derived via the magnetoconductivity measurements.[37, 48, 49]

The magnetoconductivity of a two-dimensional superconductor is mainly governed by the weak localization (WL) effect, superconducting fluctuation and the *e-e* interaction.[37, 50, 51] In the high temperature range, $T \gg T_c$, when the contributions from the Cooper pair channel are excluded, the excess magnetoconductivity for weak spin-orbit scattering can be written as[37, 48]

$$\frac{\delta\sigma(H,T)}{2\pi^2\hbar/e^2} = \delta\sigma_{2D}^{\text{WL}}(H,T) = \frac{3}{2}Y\left(\frac{H_2}{H}\right) - \frac{1}{2}Y\left(\frac{H_i}{H}\right). \tag{7}$$

Here $H_i$ is a characteristic field, which is directly related to the inelastic scattering time $\tau_{qp}$ from $\tau_{qp} = \Phi_0/4\pi D_e \mu_0 H_i$. The composite field $H_2$ describes the contribution from the spin-orbit interaction. The universal function in the two dimensional case $Y(x)$ is given by $Y(x) = \ln x + \psi(1/2 + 1/x)$ and $\psi(x)$ is the digamma function.

With the decreasing temperature, superconducting fluctuations gradually become important, which is described by the MT fluctuation theory. Thus the MT term $\delta\sigma_{2D}^{MT}(H,T)$ must be included into the excess magnetoconductivity[52-55]

$$\frac{\delta\sigma(H,T)}{2\pi^2\hbar/e^2} = \delta\sigma_{2D}^{WL}(H,T) + \delta\sigma_{2D}^{MT}(H,T) = \frac{3}{2}Y\left(\frac{H_2}{H}\right) - \frac{1}{2}Y\left(\frac{H_i}{H}\right) - \beta(T)Y\left(\frac{H_i}{H}\right). \quad (8)$$

The MT expression describes the contribution of superconducting fluctuations to the conductivity of disordered films, namely the interaction correction from the Cooper pair channel.[33, 56] The pre-factor $\beta(T)$ is strongly temperature dependent, with $\beta(T) = \pi^2/6\ln^2(T/T_C)$ at temperatures $\ln(T/T_C) \gg 1$ and $\beta(T) = \pi^2/4\ln(T/T_C)$ at temperatures $\ln(T/T_C) \ll 1$.[48, 53] The MT contribution $\delta\sigma_{2D}^{MT}(H,T)$ dominates the magnetoconductance in the temperature range where $\ln(T/T_C) < 1$ and remains accurate even far away from the superconducting fluctuation region. Unfortunately, the magnetic field range where the MT term is accurate varies strongly with temperature. For example, when the temperature approaches $T_c$, its range of validity narrows down to $H < H_i$. In order to expand the validity range to large magnetic fields for temperatures near $T_C$, the MT term has to be modified according to[31, 37, 48, 57, 58]

$$\delta\sigma_{2D}^{MT}(H,T) = -\beta(T,\delta)[Y\left(\frac{H_i}{H}\right) - Y(\frac{\widetilde{H_c}}{2H})]. \quad (9)$$

Here $\beta(T,\delta)$ can be approximated as $\pi^2/4[\ln(T/T_C) - \delta]$ at temperature $\ln(T/T_C) \ll 1$ and factor $\delta$ is the superconducting pair-breaking parameter.[33] The characteristic critical field is defined as $\mu_0\widetilde{H_c} = B_{c2}(0) \cdot \ln(T/T_C)$.[37]

The measured magnetoconductivity for the 5 nm thick $W_{0.75}Si_{0.25}$ film at different temperatures is shown in Fig. 7(a). The red curves are the fitting results based on expression (7), (8) and (9). Our magnetoconductivity data can be well described by the combination of WL effect and MT superconducting fluctuation. The extracted characteristic field $H_i$ is shown in Fig. 7(b). Using the best fitting values of $H_i$, we computed the inelastic scattering time $\tau_{qp}$ at each temperature. The results are shown in Fig. 8.

As we have discussed above, there are three main channels for the inelastic scattering. In the two-dimensional case which is appropriate for our films, the reciprocal *e-e* scattering time is $1/\tau_{e-e} = (k_B T/\hbar) \cdot [R_s/(2\pi\hbar/e^2)] \cdot \ln(\pi\hbar/R_s e^2)$,[59, 60] while the scattering rate due to *e-ph* interaction is $1/\tau_{e-ph} \propto C_1 \cdot (T/T_c)^n$ where *n* may depend on the degree of disorder and $C_1$ is a fitting parameter.[43] Superconducting fluctuations contribute to the scattering with the rate $1/\tau_{e-fl} = (k_B T/\hbar) \cdot [R_s/(2\pi\hbar/e^2)] \cdot [2\ln 2/(\ln(T/T_c) + b)]$. The exact expression for *b* can be found in reference 61. Using $C_2$ as another fitting parameter everywhere instead of $\pi\hbar/R_s e^2$, we fit the total scattering rate as the sum of the rates from these three scattering channels, $1/\tau_{qp} = 1/\tau_{e-e} + 1/\tau_{e-ph} + 1/\tau_{e-fl}$. We did not attempt to fit separately data sets for each stoichiometry, but use all available data points for the single fit. This is justified because the values of $\tau_{qp}$ are close for all samples except for temperatures in the vicinity of $T_c$. The best fit is shown in the inset in Fig. 8. For the data above $T_c$, we found out that fitting was only possible with *n* = 3 that justifies the clean limit for the *e-ph* interaction in our films. From the best fit, the resulting scattering rates are $\tau_{e-e}$ = 47 ps, $\tau_{e-ph}$ = 66 ps and $\tau_{e-fl}$ = 4.5 ps at $T = T_c$, and $C_2 = 1.1 \cdot \pi\hbar/R_s e^2$, in very good agreement with the theoretical expectation. Close to the transition, the best-fit values of the scattering rate are very sensitive to the values of *b* and $T_c$ via the *e-fl* contribution. However, they virtually do not affect the values of the scattering rates at temperatures above the transition.

### C. Time resolving recovery of superconductivity

To measure the recovery time $\tau_R$ of the superconducting state, the microbrige was cooled to *T* = 3.2 K slightly below its transition temperature $T_c$ = 3.8 K and biased with a direct current. Excitation with the light-pulse creates an impedance change, which is translated by the bias current into the voltage transient $V(t)$ between the bow-tie pads. The transients were amplified and recorded with a sampling oscilloscope. The effective bandwidth of the readout was limited to 8 GHz by the amplifier. At bias currents less than the critical current, the DC resistance was zero and the recorded transient was bipolar. Such bipolar shape is typical when a non-equilibrium state is associated with the change in the kinetic inductance.[62] The recovery of the kinetic inductance is controlled by the gap relaxation time. In order to exclude the contribution of the kinetic inductance, the detector was driven by the bias current almost into the normal-state, as it is shown in Fig. 9(a). At currents larger than the critical current, the negative part of the transient disappeared. The decaying edge becomes exponential in time $V(t) \propto \exp(-t/\tau_{falling})$ with a characteristic time $\tau_{falling}$ which initially decreases with the increase of the bias current and saturates when the actual dc resistance approaches the normal-state resistance. Fig. 9(b) shows the measured transient $V(t)$ for the 5 nm thick $W_{0.85}Si_{0.15}$ microbridge. The microbridge was biased to the operation point with the current $I = 37.3$ μA and the voltage $V = 3.6$ mV beyond which the decay time did not vary any more. A careful analysis of the falling edge of the transient response shows that $\tau_R = 628 \approx \tau_{falling} = 630$ ps after excluding the contribution from the readout to the falling edge.[63]

Recently, Marsili et al[64] used a two-photon excitation method to deduce the hot-spot relaxation time $t_{HS}$ in WSi SNSPDs. A pair of photons was introduced with a time delay onto the nanowire meander in the superconducting state. The meander was biased at a relative current less than $I/I_c = 0.65$ to operate the detector in a two-photon excitation regime. Only if the time delay is shorter than the relaxation time of the hot-spot due to the first photon, a detection event is registered. According to this method, $t_{HS}$ is derived to be around 800 ps at $T = 0.25$ K. Moreover, the hot-spot relaxation time is strongly dependent on the bias current, operating temperature, and photon energy. The authors interpreted the results according to a quasiparticle relaxation model based on the uniform kinetic equation.[65] We therefore can view the hot-spot relaxation time as the intrinsic lifetime of the quasiparticles, and the measured $\tau_R$ can be viewed as the limitation in the normal state. In both experiments, the formation of fluctuation area and the subsequent recovery are dominated by the diffusion and recombination of quasiparticles.

IV. Discussion

A. Definition of the hot-spot size

The most widely used model for a qualitative description of the detection process in SNSPD is the hot-spot model. A hot-spot is created in the nanowire after the absorption of the incident photon and then the bias current going through the hot-spot area is expelled to the sidewalks outside the hot-spot. When the hot-spot is large enough, the current density in the sidewalks will exceed the local density of the critical current. As a result, the bias current is partly shunt around through the readout line and a voltage response is created on the readout resistance.[5] We had proposed a two-stage diffusion model to describe the formation of the hot-spot.[27] To summarize, the high-energy electron which absorbs the incident photon will continuously lose its energy by means of inelastic interactions and thus creates non-equilibrium quasiparticles. The growth of their number is controlled by the inelastic electron scattering rate $\tau_{qp}$. While relaxing to low energies, the high-energy electron will move away from the point where the photon has been absorbed. This latter process can be simplified as diffusion with the diffusivity $D_e$ that gives the probability to find the hot electron at the time $t$ after the photon is absorbed at a distance $r$ from the absorption point. Simultaneously, the created quasiparticles diffuse out of this relaxation area and recombine into Cooper pairs. The local quasiparticle density $C_{qp}(r,t)$ changes due to the diffusion and recombination with the rates $D_{qp}\nabla^2 C_{qp}(r,t)$ and $C_{qp}(r,t)/\tau_r$ respectively, which are controlled by the quasiparticle diffusivity $D_{qp}$ and the recombination time $\tau_r$. Diffusion dominates the evolution of $C_{qp}(r,t)$ in the nanowire when $t > \tau_{qp}$ (assuming that the photon is absorbed at $t = 0$), and finally all quasiparticles recombine back into Cooper pairs. As a result, by first assuming $D_{qp} = D_e = D$ and neglecting the edge effects, the density of quasiparticles $C_{qp}(r,t)$ around the photon absorption area can be analytically expressed as[27]

$$C_{qp}(r,t) = \frac{\varsigma h\nu}{\Delta} \cdot \frac{\tau_r}{\tau_r - \tau_{qp}} \left[\exp\left(-\frac{t}{\tau_r}\right) - \exp\left(-\frac{t}{\tau_{qp}}\right)\right] \times \frac{1}{4\pi Dt} \exp\left(-\frac{r^2}{4Dt}\right). \tag{10}$$

Here $\varsigma$ is the energy conversion efficiency of the incident photon and $h\nu$ is the photon energy. In a more realistic situation, $D_{qp} \neq D_e$ (the excited high energy electron is different from the depaired quasiparticles) and the temperature dependence of the superconducting parameters should be considered. In such complicated conditions, we cannot give an analytical solution to the diffusion equation and only a numerical calculation can give the time evolution of the quasiparticle distribution.[27] When the operating temperature is not too low, however, $D_{qp}$ is estimated to be of the same order of magnitude with $D_e$ (e.g. in Ref. 65, $D_{qp}$ is estimated to be $0.5D_e$ at $T > 0.5T_c$), and thus we will still use this approximation $D_{qp} = D_e$ in the following. Within this simplified model, the total number of quasiparticles, which are introduced by the absorbed photon, is obtained by integrating the quasiparticle distribution within the two dimensional film[27]

$$N_{qp}(t) = \int_0^\infty C_{qp}(r,t) \cdot 2\pi r \cdot dr = \frac{\varsigma h\nu}{\Delta} \cdot \frac{\tau_r}{\tau_r - \tau_{qp}} [\exp\left(-\frac{t}{\tau_r}\right) - \exp(-\frac{t}{\tau_{qp}})]. \quad (11)$$

Although the lifetime of quasiparticles is much longer than the thermalization time $\tau_{qp}$, for the sake of generality we determine the time scale $t_{maxHS}$ at which the total number of quasiparticles reaches the maximum number from $dN_{qp}(t)/dt = 0$ as

$$t_{maxHS} = \frac{\tau_r \tau_{qp}}{\tau_r - \tau_{qp}} \ln(\frac{\tau_r}{\tau_{qp}}). \quad (12)$$

Though the quasiparticles are continuously diffusing further away from the absorption point after $t_{maxHS}$, the total number of the quasiparticles starts to decrease, and the global superconductivity begins to recover. We therefore define the hot-spot radius at $t_{maxHS}$ as

$$R_{hs} = (Dt_{maxHS})^{1/2} = [D\frac{\tau_r \tau_{qp}}{\tau_r - \tau_{qp}} \ln(\frac{\tau_r}{\tau_{qp}})]^{1/2}. \quad (13)$$

As a consequence, the hot-spot diameter in our case amounts to 107 nm for the 5 nm thick W$_{0.85}$Si$_{0.15}$ film by taking $\tau_{qp} = 9.1$ ps (Fig. 8) and assuming $\tau_r = \tau_R = 628$ ps. Note that the quasiparticle recombination time approximately equals to the electron-phonon interaction time at $T \leq T_c$ only. At an operation temperature of 0.25 K, the recombination time will grow to a few microseconds, increasing the hot-spot size to approximately 400 nm. Anyway, even at $T \approx T_c$, the expected hot-spot size is comparable with the most commonly used nanowire width of 100 nm. Neglecting the diffusion process would result in a hard core hot-spot $\sim (D\tau_{qp})^{1/2}$ with a diameter of only 52 nm. Hence, without consideration of the diffusion, the hot-spot size is underestimated significantly. According to our simplified model, both time constants play important roles in the formation of the hot-spot although $\tau_{qp}$ is significantly shorter than $\tau_r$. The hot-spot size is determined by a diffusion-based multiplication process, i.e., is mainly dominated by the lifetime of the non-equilibrium quasiparticles. With increasing bias current and decreasing temperature, Marsili *et al*[64] found that the hot-spot relaxation time $\tau_r$ increased significantly, which in turn leads to the increase of the hot-spot size.

For the present simple two stage diffusion model we did not consider the suppression of superconductivity from the incident photon and the bias current, namely the changes of the depairing energy and the order parameter. Moreover, in order to have an even more accurate description of the dynamic process and the dependence of the hot-spot size on external parameters, the escape of phonons should also be considered.

B. Relevance to the photon detection in nanowires

Since the time constant $\tau_r$ strongly depends on the bias current, operating temperature, and incident photon energy, and the quasiparticle diffusion constant $D_{qp}$ is also temperature dependent, the hot-spot size as it is defined here is also influenced by these external parameters. When the hotspot size is not large enough to shunt the bias current into the readout, the photon detection events in the non-saturation regime of the detection efficiency-bias current curve might be attributed to an assisted detection mechanism. Due to the quasiparticle cloud within the nanowire, the density of superconducting carriers will have a smooth variation across the wire and thus the local current density will have to redistribute. As a result, the energy barrier for vortex entry will be suppressed. Recent research indicated that vortices play a very important role in the photon detection process, namely excess quasiparticles reduce the edge barrier for vortices entering the nanowire or the binding energy of vortex-antivortex pairs.[66-68] From our transport measurements we found that the magnetic penetration depth of the amorphous WSi is almost twice as large as that of NbN materials. This to some extent means that the vortex can enter more easily into the WSi nanowire through the edge than in NbN based nanowires of the same width. The vortex entry barrier for a nanowire with a width of $d \ll w \ll \Lambda$, where $\Lambda(T) = 2\lambda(T)^2/d$ is the effective magnetic penetration depth for the superconducting strips,[72] can be simplified as[27, 41]

$$G(T, I_b, x) = E_B(T, I_b) \cdot \left\{ \ln\left[\frac{2w}{\pi\xi(T)}\sin\left(\frac{\pi x}{w}\right)\right] - \frac{I_b}{I_B(T,I_b)} \cdot \frac{\pi}{w} \cdot \left[x - \frac{\xi(T)}{2}\right] \right\}. \tag{14}$$

Here $E_B(T, I_b) = \Phi_0^2/2\pi\mu_0\Lambda(T)$ is the characteristic vortex energy, $I_B(T, I_b) = \Phi_0/2\pi\Lambda(T)$ is the characteristic current, and $x$ denotes the position for vortex hopping into the nanowire. A qualitative comparison can be made here between the WSi materials and the NbN. With larger magnetic penetration depth for the WSi materials, the energy barrier for a WSi based SNSPD with the same geometry is much smaller than that of the NbN based detectors. According to the discussion above, we can draw the plausible conclusion that due to vortex-assisted detection events, the $W_xSi_{1-x}$ based detector would have a higher quantum detection efficiency in the low bias current range when compared to NbN based detectors with the same device geometry. Moreover, experiments and theoretical simulations indicated that the vortex scenario should also play an important role for the dark counts.[41, 69] In this case, the WSi based devices should show much higher dark count rates than the NbN based detectors at the same normalized temperature $T/T_c$, reduced bias current $I_b/I_c$, and with the same device geometry, which shall be tested in future experiments.

V. CONCLUSIONS

We performed detailed transport measurements for three sets of amorphous $W_xSi_{1-x}$ films with different stoichiometries, and deduced from these measurements the material parameters in the superconducting and the normal state. Comparing with NbN, which is commonly used for SNSPD fabrication, the $W_xSi_{1-x}$ material possesses larger normal-state electron diffusivity, larger magnetic penetration depth, and larger superconducting coherence length. The quasiparticle thermalization time as derived from the magnetoconductivity was found to be much larger than that of NbN materials, which is most probably due to the amorphous nature of tungsten silicide. The electron-energy relaxation time was extracted from the time-resolving measurements of the recovery of the superconducting state after femtosecond pulse excitation.

Within a two stage diffusion model, we found that the formation of a hot-spot is controlled by an initial thermalization process of the electron which absorbed the incident photon, and a subsequent diffusion and recombination of non-equilibrium quasiparticles. As a result, a hot-spot diameter of 105 nm was estimated for the 5 nm thick $W_{0.85}Si_{0.15}$ film near the transition temperature. Finally, within the vortex-assisted photon detection model, we expect a higher detection efficiency for $W_xSi_{1-x}$ based detectors for low energy photons or in the low bias current range as compared to NbN based detectors.

ACKNOWLEDGEMENT

M. Sidorova acknowledges support of the Helmholtz Research School on Security Technologies at DLR.

TABLE I. Material parameters of the 5 nm thick films. The material parameters from NbN and TaN SNSPD are also listed.

| Sample | $d$ (nm) | $w$ (μm) | $R_{ns}$ (Ω) | $T_c(0)$ (K) | $\xi(0)$ (nm) | $D_e$ (cm²/s) | $N_0(0)$ (m⁻³J⁻¹) | $\Delta(0)$ (meV) | $\lambda(0)$ (nm) |
|---|---|---|---|---|---|---|---|---|---|
| $W_{0.75}Si_{0.25}$ | 5 | 10 | 410 | 3.86 | 7.1 | 0.71 | $2.7 \times 10^{47}$ | 0.59 | 763 |
| $W_{0.75}Si_{0.25}$ | 5 | 100 | 417 | 3.88 | 7.3 | 0.70 | $2.6 \times 10^{47}$ | 0.59 | 768 |
| $W_{0.8}Si_{0.2}$ | 5 | 10 | 340 | 4.02 | 7.1 | 0.71 | $3.2 \times 10^{47}$ | 0.61 | 696 |
| $W_{0.8}Si_{0.2}$ | 5 | 100 | 357 | 4.08 | 7.0 | 0.70 | $3.1 \times 10^{47}$ | 0.61 | 681 |
| $W_{0.85}Si_{0.15}$ | 5 | 10 | 326 | 3.83 | 7.3 | 0.73 | $4.1 \times 10^{47}$ | 0.58 | 735 |
| $W_{0.85}Si_{0.15}$ | 5 | 100 | 350 | 3.85 | 7.4 | 0.75 | $3.8 \times 10^{47}$ | 0.59 | 706 |
| NbN[41] | 6 | 0.053 | 445 | 12.73 | 4.0 | 0.49 | $3.6 \times 10^{47}$ | 2.30 | 404 |
| NbN[16] | 6 | 0.08 | 380 | 13.0 | 4.3 | 0.50 | $5.1 \times 10^{47}$ | 1.98 | 440 |
| TaN[16] | 3.9 | 0.126 | 380 | 9.30 | 5.0 | 0.60 | $4.4 \times 10^{47}$ | 1.24 | 490 |

In references 16 and 41, correction of $\xi(0)$ is adopted since the real $B_{c2}(0)$ is smaller than the linearly extrapolated $B_{c2}(0)$. With this correction factor, the $\xi(0)$ of WSi will be slightly larger than the values listed above.

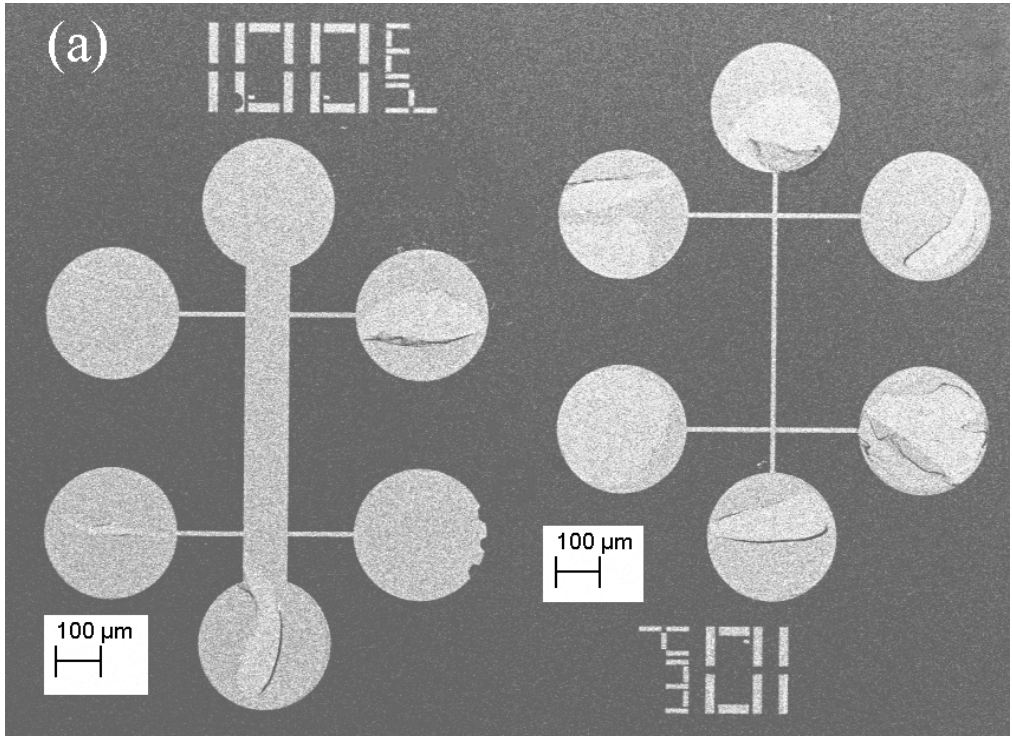

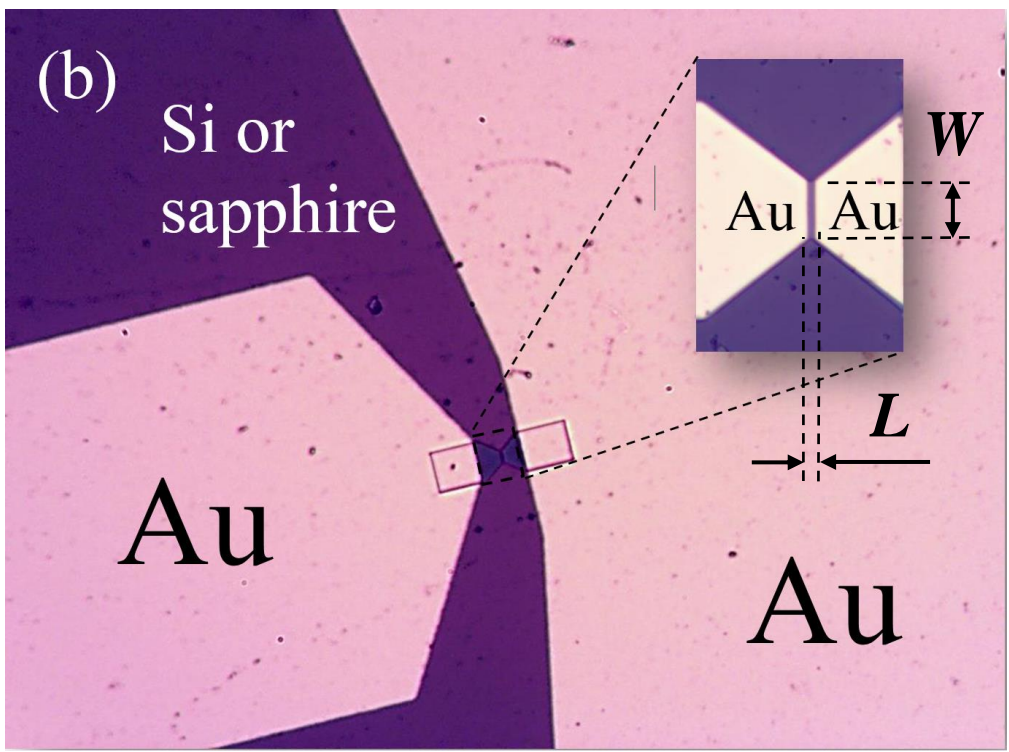

FIG. 1. SEM images of the specimens. (a) The microstrip used for transport measurement. The calculations in this paper are based on the measured strip geometries. (b) The bow-tie structure used for the $\tau_R$ measurement. The inset shows the enlarged sensitive area and the WSi microbridge located between the two gold pads.

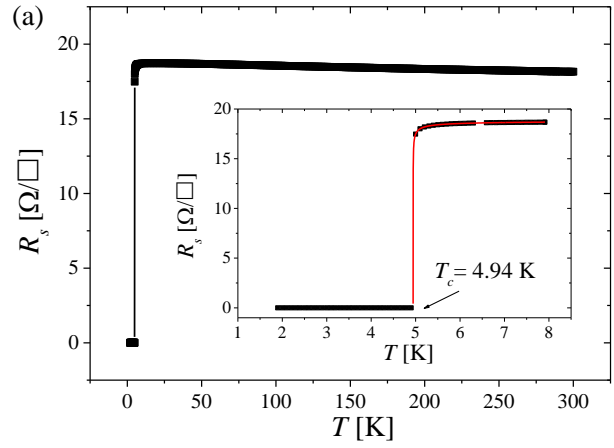

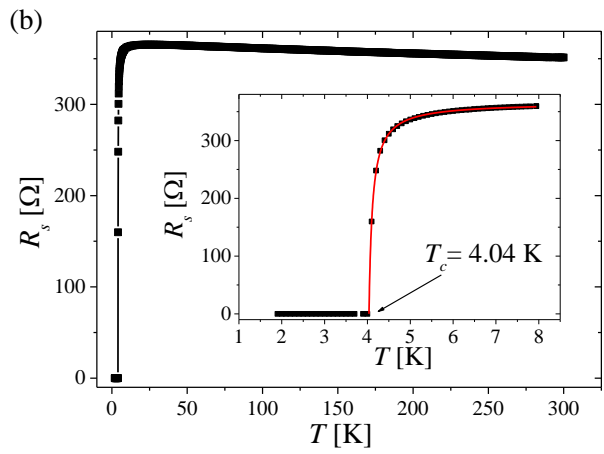

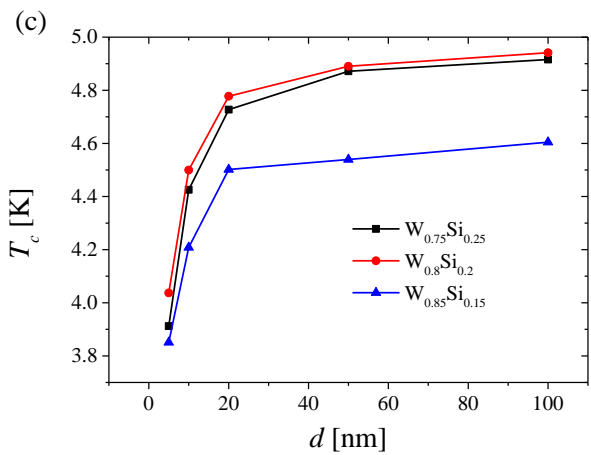

FIG. 2. (a) Square resistance for the 100 nm thick and 100 μm wide $W_{0.8}Si_{0.2}$ strip as a function of temperature. The superconducting transition is fitted with Eq. (1). (b) Temperature dependence of the square resistance from the 5 nm thick and 100 μm wide $W_{0.8}Si_{0.2}$ strip. The superconducting transition is fitted with Eq. (2). (c) The mean-field critical temperatures as functions of the film thickness.

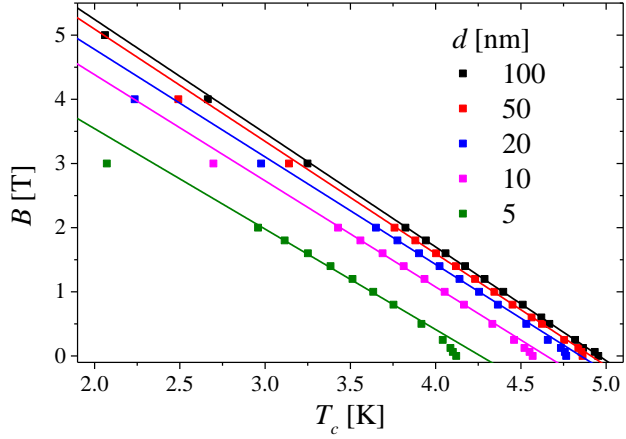

FIG. 3. The critical magnetic field at different temperatures for a series of 100 μm wide $W_{0.8}Si_{0.2}$ strips. Through linear fitting of these temperature dependences we extracted the zero-temperature critical magnetic field $B_{c2}(0)$.

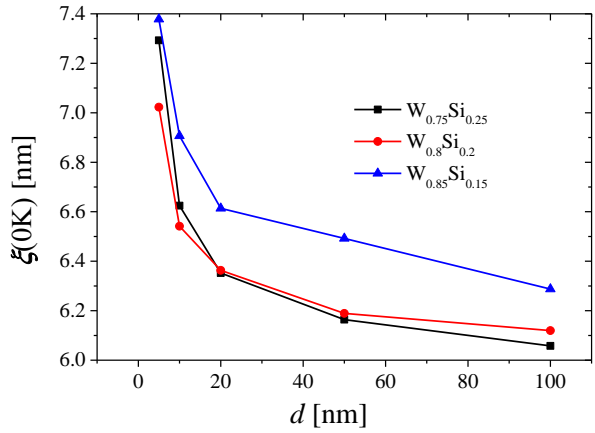

FIG. 4. The GL coherence length at zero temperature $\xi(0)$ as a function of film thickness.

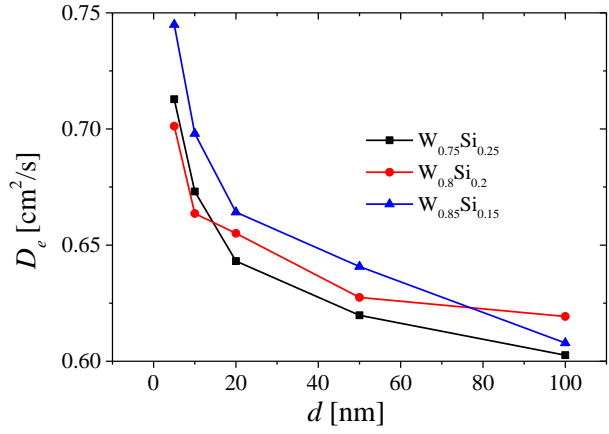

FIG. 5. Thickness dependence of the diffusion constant of the electrons in the normal-conducting state.

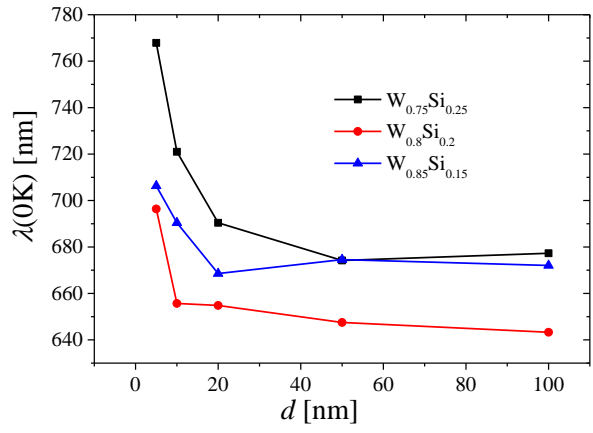

FIG. 6. The magnetic penetration depth $\lambda(0)$ as a function of film thickness.

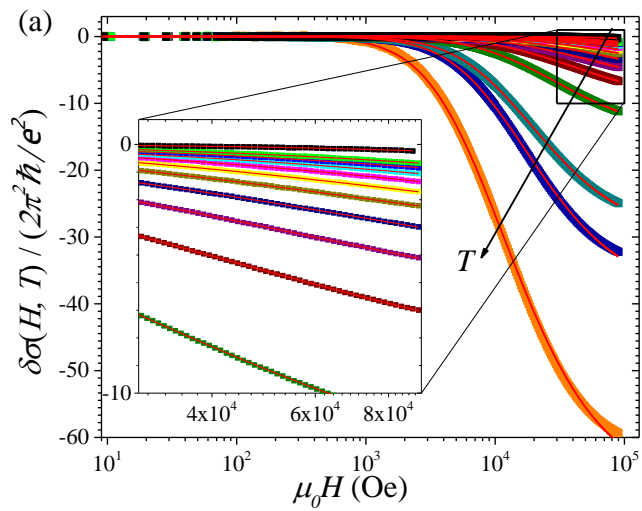

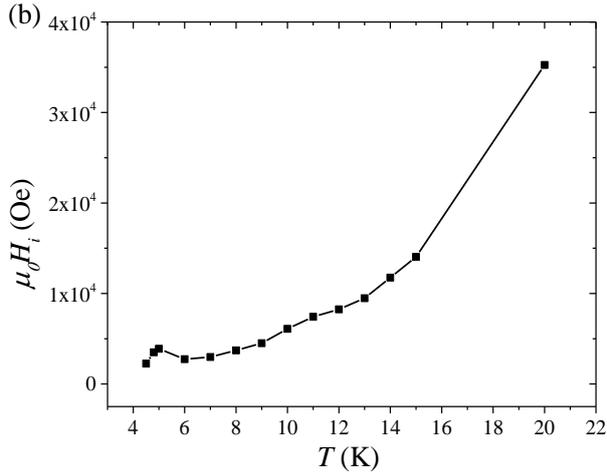

FIG. 7. (a) The excess magnetoconductivity $\delta\sigma(H,T)/(2\pi^2\hbar/e^2)$ vs applied magnetic field at temperatures $T = 20$ K (black), 15 K (green), 14 K (blue), 13 K (cyan, 12 K (magenta), 11 K (yellow), 10 K (dark yellow), 9 K (navy), 8K (purple), 7 K (wine), 6 K (olive), 5 K (dark cyan), 4.8 K (Royal), 4.5 K (orange). The black arrow indicates the decreasing temperature. The red curves are fits using the Eqs. (7), (8) and (9). Inset: magnification of the detailed MC data in the high temperature and high magnetic field range. (b) The characteristic magnetic field extracted from the fitting procedure as a function of temperature.

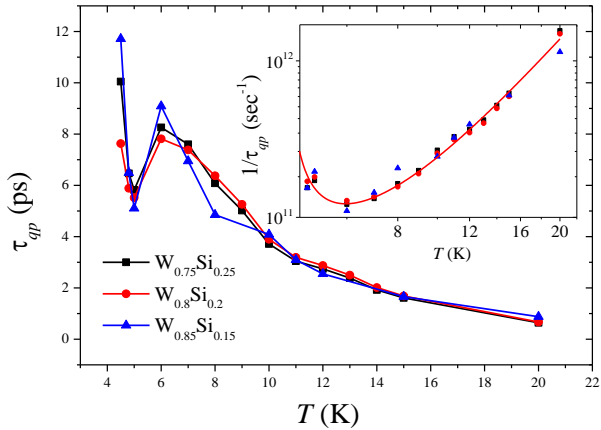

FIG. 8. The inelastic scattering time calculated from the characteristic field $H_i$ for 5 nm thick films. Inset: the inelastic scattering rate $1/\tau_{qp}$ vs temperature for $W_{0.75}Si_{0.25}$ (black), $W_{0.8}Si_{0.2}$ (red), and $W_{0.85}Si_{0.15}$ (blue) stoichiometries. The solid curve shows the best fit according to $1/\tau_{qp} = 1/\tau_{e-e} + 1/\tau_{e-ph} + 1/\tau_{e-fl}$.

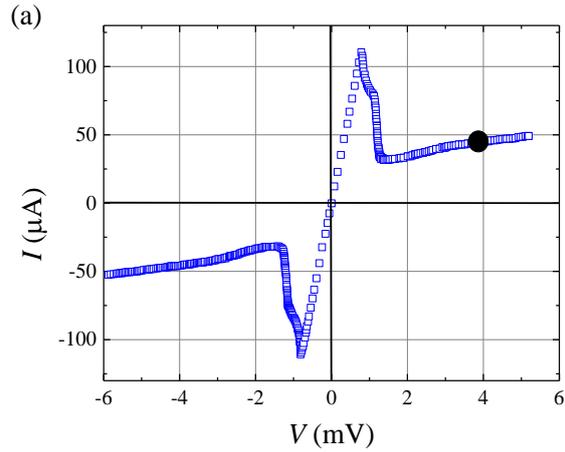

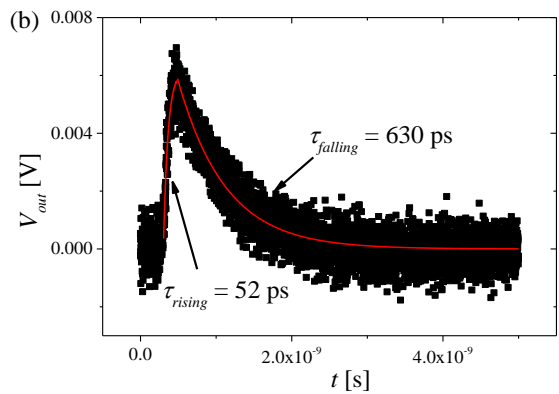

FIG. 9. (a) The *I-V* curve of the $W_{0.85}Si_{0.15}$ microbridge measured at $T = 3.2$ K. The critical current for the device is around 119 µA and the series resistance of the bias circuit is 14 Ω. The filled dot shows the regime where the voltage response was measured. (b) The voltage response vs time. The rising and the falling edges of $V_{out}(t)$ were fitted separately to exponential functions with time constants $\tau_{rising}$ and $\tau_{falling}$, respectively.